\documentclass[11pt,a4paper]{article}
\usepackage{amsmath,amsthm,amssymb}
\usepackage{geometry}
\usepackage{hyperref} % optional, but OK
\usepackage{epsfig}
\usepackage{graphicx,color}
\usepackage{float}
\theoremstyle{plain}
\newtheorem{theorem}{Theorem}[section]

\theoremstyle{definition}

\numberwithin{equation}{section}

\begin{document}

\title{Numerical Methods for a 2D ``Bad'' Boussinesq Equation: RK4, Strang Splitting, and High-frequency Fourier Modes}

\author{Arief Anbiya \\ Independent Researcher, Indonesia \\ \texttt{ariefanbiya@gmail.com}}
\date{\today}
\maketitle

\begin{abstract}
Numerical methods for a two-dimensional ``bad'' Boussinesq equation: $u_{tt} = u_{xx} + u_{xxxx} + u_{yy} - 3 (u^{2})_{xx}$ are presented with good accuracy. The methods mainly depend on pseudo-spectral Fourier with a trimming of carefully chosen high-frequency Fourier modes. One method also relies on Runge-Kutta fourth order (RK4), and another method relies on Strang operator splitting. Before implementing the two methods, we analyze using Fourier series the linearized version of the equation by removing the nonlinear term $3(u^{2})_{xx}$, and found that a particular bound or condition needs to be satisfied to avoid blow-up solution. We found that high-frequency Fourier modes that do not satisfy the condition must be excluded from the Fourier solution. We then apply this condition to the numerical methods for solving the nonlinear Boussinesq equation and found that including only the Fourier modes that satisfy the condition gives stable numerical solution with good accuracy up to $t=100$. Including even just a few number of Fourier modes that violate the condition can result in a blow-up solution as early as $t=23.5$. The accuracy of the method is measured by computing the $L^{\infty}$ error against a soliton exact solution. The errors resulting from RK4 and Strang splitting numerical simulations differ slightly for small $\triangle t$, while there is a noticeable decrease in performance for the Strang splitting simulation as $\triangle t$ increases. Using our numerical methods, we also display a simulation with Dirichlet boundary condition to account for wave reflections.
\end{abstract}

\section{INTRODUCTION} %This is standard; but you may have better titles for the
%sections of your paper.

\noindent 
The following one-dimensional classical Boussinesq equation was derived by J. Boussinesq \cite{j_boussinesq} to model the propagation of dispersive long water waves with small amplitude:
\begin{equation}\label{1d_bbe}
u_{tt} = u_{xx} + u_{xxxx} + (u^{2})_{xx}.
\end{equation} 
Although initially it was not explicitly referred to as ``bad'',  it has now often been coined as ``bad'' Boussinesq equation due to that its linearized version, $u_{tt} = u_{xx} + u_{xxxx}$, is not a well-posed problem in the Hadamard sense \cite{hadamard_wellposed} for bounded domain. The ill-posedness can be revealed by using Fourier series on the linearized equation: for high enough frequencies, the Fourier coefficients can grow exponentially large, which means a small change of the initial condition can give a large change in the solution. Also, the ill-posedness persists in the nonlinear level:  there exists a class of solutions of (\ref{1d_bbe}) that `blow up' \cite{charlier_blowup,charlier_blowup2}. More recent analysis by the same authors including on its' asymptotic solutions can be seen in \cite{charlier_analysis_1},\cite{charlier_analysis_2},\cite{charlier_analysis_3},\cite{charlier_analysis_4}. On the other hand, if we change the sign of the $u_{xxxx}$ term to negative, we have the `good' Boussinesq equation which is linearly well-posed and not as intricate as (\ref{1d_bbe}) to be solved numerically. Papers on numerical solutions for the `good' Boussinesq equation can be seen in \cite{fourth_order_good_boussinesq} which uses a fourth order finite difference method, and in \cite{strang_splitting_good_boussinesq} that uses Strang operator splitting with high accuracy. Although the `good' Boussinesq equation has applications and it is linearly well-posed and easier to solve, 

\begin{figure}[H]
\centering
\includegraphics[width=14cm]{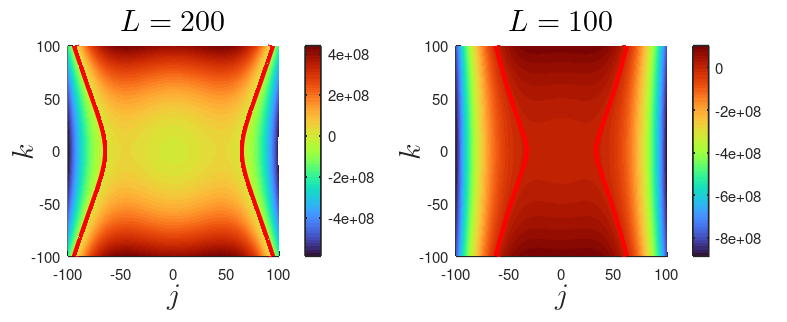}
\caption{Comparison of the function $F(j,k) = L^{2}k^{2} +  L^{2}j^{2} - \pi^{2} j^{4}$ when $L=100$ and $L=200$. The red lines are where $F(j,k)=0$.}
\label{fig:trimming_comparison}
\end{figure}

\noindent it does not model the same physical wave phenomenon as the ``bad'' Boussinesq equation. In \cite{charlier_numerics}, a numerical scheme for (\ref{1d_bbe}) is presented with good results. The scheme uses Fourier spectral method by trimming the high-frequency Fourier modes that have frequencies exceeding a bound. The bound is carefully chosen such that the solution of the linearized equation, via Fourier spectral method, does not blow-up or grow exponentially. The simple bound also reveals that wider domain, in the case of 1-D, allows more high-frequency modes to be included in the Fourier solution, which is consistent for modelling long waves that obviously need wide domain.

Inspired by \cite{charlier_numerics}, we want to adapt their method on a two-dimensional Boussinesq equation. A two-dimensional version was derived in 1996 as
\begin{equation}\label{2d_bbe}
u_{tt} = u_{xx} + u_{xxxx} + u_{yy} - 3 (u^{2})_{xx}.
\end{equation}
The nonlinear equation (\ref{2d_bbe}) combines the two-way propagation of the classical Boussinesq equation in one-dimensional space with a weak dependence on the second spatial variable $y$ \cite{rs_johnson}. As you will see in section \ref{linearized_solution}, the linearized version of (\ref{2d_bbe}) can also be shown to be ill-posed using Fourier spectral method: it has high-frequency Fourier modes that can grow exponentially in time. We then derive the condition that must be satisfied to avoid blow-up solution. First, we derive the `trimming' condition for the case of square domain (\ref{stable_condition}), before also generalize to rectangular domain (\ref{stable_condition_2}). Subsequently, we use this condition and apply Fourier spectral method \cite{pseudo_spectral} to solve the two-dimensional ``bad'' Boussinesq equation numerically. Similar to \cite{charlier_numerics}, our numerical scheme is implemented by first rewrite (\ref{2d_bbe}) as a system (\ref{system}). We then view the approximate solution for $u$ in terms of two-dimensional Fourier series. After substituting with Fourier series, we then must solve a system of ordinary differential equations in the frequency space and ``trimmed'' the Fourier modes with high enough frequencies that exceed our bound. We then compare our results by computing the numerical error against the exact solution presented in \cite{exact_barrera} in the form of soliton. We then also solve (\ref{2d_bbe}) numerically with a Dirichlet boundary condition to simulate wave reflections. The numerical scheme is found to provide stable and accurate enough approximation. The numerical results in section \ref{sec:numerical_scheme} show that the $L^{\infty}$ errors between the numerical and exact solutions are small enough: less than 3\% of the absolute initial peak $\max(|U(x,y,0)|)$ for $t \in [0,40]$. 

The solving of the system of PDEs involves two numerical methods: the Runge-Kutta fourth order (for solving ODEs in frequency space) and the Strang operator splitting (with exact solution in frequency space), both are paired with the `trimming' condition (\ref{stable_condition_2}). The Strang splitting is a method for solving a differential equation by first separating between the linear and nonlinear parts, before combining the solutions of each individual part in a particular way. The method has been proven to have second order convergence on the nonlinear Korteweg de Vries (KdV) equation \cite{tao_kdv} and has been tested numerically on the `good' Boussinesq equation

\begin{figure}[H]
\centering
\includegraphics[width=12cm]{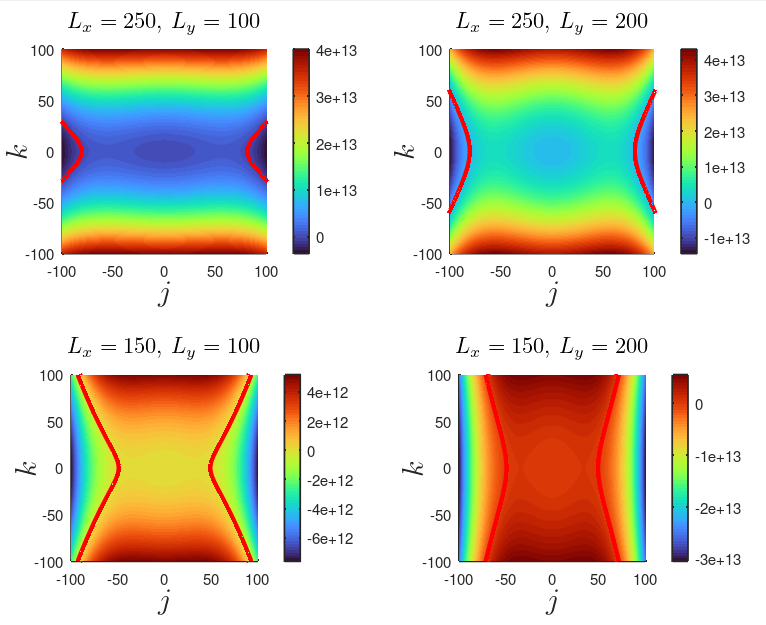}
\caption{Comparison of the function $F_{2}(j,k) = L_{x}^{4} k^{2} + L_{y}^{2}\left( L_{x}^{2}j^{2} - \pi^{2}j^{4} \right)$ with four combinations of $L_{x}$ and $L_{y}$. The red lines are where $F_{2}(j,k)=0$.}
\label{fig:trimming_comparison2}
\end{figure}

\noindent with good accuracy \cite{strang_splitting_good_boussinesq}. In this paper, we found that the method can also be used to approximate the two-dimensional Boussinesq equation (\ref{2d_bbe}) by first writing it as a first order system. The numerical simulations are implemented in Octave software \cite{octave}, the code for the numerical simulations are publicly available in the author's Github repository \url{https://github.com/anbarief/Boussinesq2D_Fourier/tree/main}.

The main contributions of this paper can be summarized as follows:
\begin{itemize}
\item We extend and improvise from the numerical method presented in \cite{charlier_numerics} for 1D ``bad'' Boussinesq equation to a numerical method for a 2D ``bad'' Boussinesq equation, by trimming some high-frequency Fourier modes in the pseudo-spectral implementation. We derive the trimming condition by analyzing the Fourier series solution of the linearized equation (\ref{2d_bbe_linear}). Although the approach is similar as \cite{charlier_numerics}, the obtained trimming condition (\ref{stable_condition}) in this paper is different and more nuanced.
\item In the pseudo-spectral numerical scheme, we compare two methods for solving the system of ODEs in the frequency space: Runge-Kutta fourth order and Strang operator splitting. Before using both methods, we first rewrite the nonlinear equation (\ref{2d_bbe}) as a first-order system. This is a mild improvisation of the work in \cite{strang_splitting_good_boussinesq}, in which they also similarly rewrite the ``good'' Boussinesq equation as a first order system. 
\item Although the trimming condition is derived from the linearized equation (\ref{2d_bbe_linear}), it has significance when used in the nonlinear equation (\ref{2d_bbe}). We found that the numerical solution of (\ref{2d_bbe}) is stable up to $t=100$ (see Fig \ref{fig:long_sim_3d}), if we trim the high frequency modes that violate the stability condition (\ref{stable_condition}). On the other hand, a mild violation of (\ref{stable_condition}) result in a blow-up as early as $t=23.5$ (see Fig \ref{fig:blowup_example}). This may indicate that the linear term $u_{xxxx}$ contributes more on the blow-up behavior, rather than the nonlinear term $-3(u^{2})_{xx}$.
\end{itemize}

\section{LINEARIZED SOLUTION} \label{linearized_solution}%You may have better titles for the sections of your paper.
\noindent We first consider the linearized version of (\ref{2d_bbe}) that is attained by removing the term $3(u^{2})_{xx}$ on a square domain:
\begin{equation}\label{2d_bbe_linear}
u_{tt} = u_{xx} + u_{xxxx} + u_{yy}, \:\:\: (x,y) \in \Omega = (-L,L)^{2}, \:\:\: t > 0,
\end{equation}
with initial conditions $u(x,y,0)=u_{0}(x,y)$ and $u_{t}(x,y,0)=v_{0}(x,y)$. We consider the $2L$-periodic solution of (\ref{2d_bbe_linear}) in terms of two-dimensional Fourier series on a square domain $(-L,L)^{2}$. Since the solution is assumed to be periodic by Fourier, then the equation is not conserving mass except when the condition $\int_{\Omega} u_{t}(x,y,0) \: d\Omega = 0$ is met, so we also assume $v_{0}(x,y)$ satisfies this. To see this, let the total mass at time $t$ be $M(t) = \int_{\Omega} u(x,y,t) \: d\Omega$, since Fourier-periodicity of $u(x,y,t)$ implies periodicity of its derivatives, then we have
\begin{align*}
\int_{-L}^{L} \left( \int_{-L}^{L} u_{xx} \: dx \right) \: dy &= \int_{-L}^{L} \left( u_{x}(L,y,t) -  u_{x}(-L,y,t)   \right) \: dy = 0\\
\int_{-L}^{L} \left( \int_{-L}^{L} u_{yy} \: dy \right) \: dx &= \int_{-L}^{L} \left( u_{y}(x,L,t) -  u_{y}(x,-L,t)   \right) \: dx = 0\\
\int_{-L}^{L} \left( \int_{-L}^{L} u_{xxxx} \: dx \right) \: dy &= \int_{-L}^{L} \left( u_{xxx}(L,y,t) -  u_{xxx}(-L,y,t)   \right) \: dy = 0.
\end{align*}
By equation (\ref{2d_bbe_linear}), this implies $\int_{\Omega} u_{tt} \: d \Omega = M''(t) = 0$, so that $M'(t)$ is constant. Thus, condition $M'(0)=\int_{\Omega} u_{t}(x,y,0) \: d\Omega = 0$ guarantees $M'(t)=0$ for all $t$, or $M(t)$ always constant.  Next, the solution can be written as follows
\begin{align*}
u(x,y,t) = \sum _{k \in \mathbb{Z} } \sum_{j \in \mathbb{Z}} \widehat{u}_{j,k}(t)e^{i 2 \pi (\omega_{j}x + \omega_{k}y)},
\end{align*}
where $\omega_{j}=j/2L, \: \omega_{k}=k/2L$. By replacing $u$ in the linear equation (\ref{2d_bbe_linear}) with Fourier series, we have the following
\small \begin{align*} 
\sum _{k \in \mathbb{Z}} \sum_{j \in \mathbb{Z}} \widehat{u}_{j,k}''(t)e^{i 2 \pi (\omega_{j}x + \omega_{k}y)} &= \sum _{k \in \mathbb{Z}} \sum_{j \in \mathbb{Z}} (i 2 \pi \omega_{j})^{2} \widehat{u}_{j,k}(t)e^{i 2 \pi (\omega_{j}x + \omega_{k}y)} \\
&+ \sum _{k \in \mathbb{Z}} \sum_{j \in \mathbb{Z}} (i 2 \pi \omega_{j})^{4} \widehat{u}_{j,k}(t)e^{i 2 \pi (\omega_{j}x + \omega_{k}y)} \\
&+ \sum _{k \in \mathbb{Z}} \sum_{j \in \mathbb{Z}} (i 2 \pi \omega_{k})^{2} \widehat{u}_{j,k}(t)e^{i 2 \pi (\omega_{j}x + \omega_{k}y)}.
\end{align*} \normalsize
Solving the above is equivalent as solving the following system
\begin{equation}\label{2d_bbe_linear_frequency_domain}
\widehat{u}_{j,k}'' = \left[ ( 2 \pi \omega_{j})^{4} - ( 2 \pi \omega_{k})^{2} -  (2 \pi \omega_{j})^{2} \right] \widehat{u}_{j,k}  , \:\:\: \forall (j,k) \in \mathbb{Z}^{2}, \:\:\: t > 0.
\end{equation}
The solution of each equation above depends on $( 2 \pi \omega_{j})^{4} - ( 2 \pi \omega_{k})^{2} -  (2 \pi \omega_{j})^{2}$, which we can categorize into three cases. Let $\lambda_{j,k} = ( 2 \pi \omega_{j})^{4} - ( 2 \pi \omega_{k})^{2} -  (2 \pi \omega_{j})^{2}$ for each pair $(j,k)$, then we have the following
\begin{equation} \label{uhat_exact_solution}
\widehat{u}_{j,k} = 
\begin{cases} 
\widehat{u}'_{j,k}(0)t + \widehat{u}_{j,k}(0), \:\:\: \lambda_{j,k} = 0 \\
  \frac{\widehat{u}_{j,k}(0)+ \lambda_{j,k}^{-1/2}\widehat{u}_{j,k}'(0)}{2} e^{\sqrt{\lambda_{j,k}} t} + \left( \widehat{u}_{j,k}(0) - \frac{\widehat{u}_{j,k}(0)+\lambda_{j,k}^{-1/2}\widehat{u}_{j,k}'(0)}{2} \right) e^{-\sqrt{\lambda_{j,k}} t}, \:\:\: \lambda_{j,k} > 0 \\
\frac{\widehat{u}_{j,k}(0)+ \frac{\widehat{u}_{j,k}'(0)}{ i\sqrt{-\lambda_{j,k}}}}{2} e^{i\sqrt{-\lambda_{j,k}} t} +  \left( \widehat{u}_{j,k}(0) - \frac{\widehat{u}_{j,k}(0)+\frac{\widehat{u}_{j,k}'(0)}{ i\sqrt{-\lambda_{j,k}}}}{2} \right)e^{-i\sqrt{-\lambda_{j,k}} t}  , \:\:\: \lambda_{j,k} < 0. 
\end{cases}
\end{equation}
Due to the term $e^{\sqrt{\lambda_{j,k}}t}$ for  the case $\lambda_{j,k} > 0$ in (\ref{uhat_exact_solution}), we can see that some of the Fourier coefficients can grow exponentially large as $t$ increases. Therefore, the solution $u$ can be unstable and blow-up solutions may occur if we include certain Fourier modes. The case $\lambda_{j,k}=0$ may also generate Fourier mode that keeps increasing. However, it can happen only when $j=k=0$ (provided we use integer $L$), and $\widehat{u}_{0,0}(t)$ will be constant due to periodicity and mass conservation: 
$$ \text{Constant} = \int_{\Omega} u \: d \Omega = \int_{\Omega} \widehat{u}_{0,0}(t) d\Omega + \underbrace{\int_{\Omega} \sum _{(j,k) \in \mathbb{Z}^{2} - {(0,0)}} \widehat{u}_{j,k}(t)e^{i 2 \pi (\omega_{j}x + \omega_{k}y)} \: d \Omega}_{0} =  \widehat{u}_{0,0}(t) (2L)^{2} $$ 
Since the case $\lambda_{j,k} \le 0$ gives stable solution, we attempt to neglect the Fourier modes that result in $\lambda_{j,k} > 0$. The condition for $\lambda_{j,k} \le 0$ can be derived as
\begin{align*}
( 2 \pi \omega_{j})^{4} - ( 2 \pi \omega_{k})^{2} -  (2 \pi \omega_{j})^{2} &\le 0 \\
( 2 \pi (j/2L))^{4} - ( 2 \pi (k/2L))^{2} -  (2 \pi (j/2L))^{2} &\le 0 \\
(  \pi j/L)^{4} - ( \pi k/L)^{2} -  ( \pi j/L)^{2} &\le 0 \\ 
  (\pi j)^{4}/L^{2} - ( \pi k)^{2} -  ( \pi j)^{2} &\le 0 \\
  (\pi j)^{4}/L^{2} &\le ( \pi k)^{2} +  ( \pi j)^{2},
\end{align*}
which then gives us
\begin{equation}\label{stable_condition}
     0 \le L^{2}k^{2} +  L^{2}j^{2} - \pi^{2} j^{4}.
\end{equation}
From (\ref{stable_condition}) we see that using a small domain (small $L$) will limit the number of `stable' Fourier modes that we can use: the number of combinations $(j,k)$ that we can use is limited. Using wide domain (large $L$) should give better accuracy due to more `stable' Fourier modes (see Figure \ref{fig:trimming_comparison}). We want the following function  
$$F(j,k) = L^{2}k^{2} +  L^{2}j^{2} - \pi^{2} j^{4} $$
to be non-negative. From Figure \ref{fig:trimming_comparison}, we can see that this is achieved in between the red lines. Note also that we can have high number for $k$, so we can have high frequencies without limit for the Fourier modes in $y$-direction. Note also that this inhibiting property does not violate the water wave model, since the assumptions for long wave imply that the domain is infinitely large ($L \rightarrow \infty$), in which case the Fourier series solution can be fully used. However, in the numerical approximation, we need to exclude high-frequency modes that do not follow (\ref{stable_condition}), and this can be a good approximation since long waves rely more on low-frequency Fourier modes. We will use the condition (\ref{stable_condition}) for the nonlinear Boussinesq equation (\ref{2d_bbe}), even though we derived this from the linearized equation. We will see that using this condition gives stable nonlinear numerical solution. Violating the condition, even only a little, result in a blow-up solution (see Figure \ref{fig:blowup_example}). Also note that, if we assume $u^{2}$ as a $2L$-periodic Fourier series, then condition $M'(0)=0$ for mass conservation also applies in the nonlinear equation, since $\int_{\Omega} (u^{2})_{xx} \: d\Omega = 0$.

\section{RECTANGULAR DOMAIN}
\noindent We can extend the method for a rectangular domain. The linearized equation on a rectangular domain is
\begin{equation}\label{2d_bbe_linear_rectangular}
u_{tt} = u_{xx} + u_{xxxx} + u_{yy}, \:\:\: (x,y) \in \Omega = (-L_{x},L_{x}) \times (-L_{y},L_{y}), \:\:\: t > 0,
\end{equation}
with initial conditions $u(x,y,0)=u_{0}(x,y)$ and $u_{t}(x,y,0)=v_{0}(x,y)$. We consider the periodic solution of (\ref{2d_bbe_linear_rectangular}) in terms of two-dimensional Fourier series. By the same approach for the square domain, it is not difficult to see that $M'(0)=0$ guarantees mass conservation. Plugging the Fourier series into the equation and then solving the ODE in frequency space will give us the same result as (\ref{uhat_exact_solution}), but with frequencies defined differently as $\omega_{j}=j/(2L_{x}), \: \omega_{k}=k/(2L_{y})$. The condition for $\lambda_{j,k} \le 0$ can then be derived as
\begin{align*}
( 2 \pi \omega_{j})^{4} - ( 2 \pi \omega_{k})^{2} -  (2 \pi \omega_{j})^{2} &\le 0 \\
( 2 \pi (j/2L_{x}))^{4} - ( 2 \pi (k/2L_{y}))^{2} -  (2 \pi (j/2L_{x}))^{2} &\le 0 \\
(  \pi j/L_{x})^{4} - ( \pi k/L_{y})^{2} -  ( \pi j/L_{x})^{2} &\le 0 \\ 
 \pi^{2}( j/L_{x})^{4} - (  k/L_{y})^{2} -  (  j/L_{x})^{2} &\le 0 \\
L_{y}^{2}\pi^{2}j^{4} - L_{x}^{4} k^{2} -  (L_{x}L_{y})^{2}j^{2} &\le 0, 
\end{align*}
which then gives us
\begin{equation}\label{stable_condition_2}
0 \le L_{x}^{4} k^{2} + L_{y}^{2}\left( L_{x}^{2}j^{2} - \pi^{2}j^{4} \right).
\end{equation}
We want the following bivariate function to be non-negative:
$$F_{2}(j,k) = L_{x}^{4} k^{2} + L_{y}^{2}\left( L_{x}^{2}j^{2} - \pi^{2}j^{4} \right). $$
The plots of $F_{2}(j,k)$ are shown for several combinations of $L_{x},L_{y}$ in Figure \ref{fig:trimming_comparison2}.

From (\ref{stable_condition_2}), we can see that the first term on the right hand side, $L_{x}^{4}k^{2}$, will always be non-negative. Also, $k$ only appears in this first term as a factor $k^{2}$. Thus, large values of $|k|$ will always be welcomed for stability. However, the factor $L_{x}^{2}j^{2} - \pi^{2}j^{4}$ of the second term can be negative, depending on how large $|j|$ is relative to $L_{x}$. Thus, for a large enough $|j|$, we need to combine them with large values of $|k|$ for stability, thus sacrificing low values of $|k|$.
Once the second term becomes negative, the absolute magnitude of this term is amplified by $L_{y}^{2}$. Thus, if the value $L_{x}$ stays the same, increasing $L_{y}$ will sacrifice more low values of $|k|$. In this regard, we present Theorem \ref{theorem:1} on the stability condition (\ref{stable_condition_2}). By Theorem \ref{theorem:1}, we can conclude that there is no upper bound of $|k|$, but there is a lower bound that depends on $j$. Also, by Theorem \ref{theorem:1}, we can conclude that $|j|$ has only an upper bound that depends on $k$.

\begin{theorem}\label{theorem:1}
Given $L_{x},L_{y} >0$, the ordered pair of values $(j,k)$ that satisfy (\ref{stable_condition_2}) must follow the following rules:
\begin{enumerate}
\item $|k|$ is bounded below. The lower bound depends on $j$:
\begin{align} 
\sqrt{\frac{L_{y}^{2}}{L_{x}^{4}}\left[   \pi^{2}j^{4} - L_{x}^{2}j^{2}\right]} \le  |k|, \:\:\: &\text{if } \frac{L_{y}^{2}}{L_{x}^{4}}\left[   \pi^{2}j^{4} - L_{x}^{2}j^{2}\right] \ge 0  \label{k_bound} \\
k \in \mathbb{R}, \:\:\: &\text{if } \frac{L_{y}^{2}}{L_{x}^{4}}\left[   \pi^{2}j^{4} - L_{x}^{2}j^{2}\right] < 0, \nonumber
\end{align}
\item $|j|$ is not bounded below by any positive number, but is bounded above. The upper bound depends only on $k$:
\begin{equation}
    |j| \le L_{x}\sqrt{\frac{1}{2 \pi^{2}} + 
\frac{\sqrt{ L_{y}^{2} +  4(k \pi)^{2}}}{2 \pi^{2}L_{y}}}. \label{j_bound}
\end{equation}
\end{enumerate}
\end{theorem}
\begin{proof}
To prove the lower bound of $|k|$, we simply rewrite (\ref{stable_condition_2}) as $\frac{L_{y}^{2}}{L_{x}^{4}}\left[ \pi^{2}j^{4} - L_{x}^{2}j^{2}  \right] \le  k^{2}$, the result is then obvious. To prove the upper bound of $|j|$, let $z = j^{2}$, then we must have
\begin{align*}
 L_{y}^{2}\left[ \pi^{2}z^{2} - L_{x}^{2}z\right] &\le L_{x}^{4} k^{2}   \\
 \pi^{2}z^{2} - L_{x}^{2}z  - \frac{L_{x}^{4}k^{2}}{L_{y}^{2}}   &\le 0
 \end{align*}
The determinant of the above quadratic form is 
$$ D = L_{x}^{4} \left( \frac{L_{y}^{2} +  4(k \pi)^{2}}{L_{y}^{2}} \right) > 0 $$
The roots are
$$z_{1,2} = \frac{1}{2 \pi^{2}} \left( L_{x}^{2} \pm \frac{L_{x}^{2}}{L_{y}}\sqrt{ L_{y}^{2} +  4(k \pi)^{2}} \right) $$
It is easy to see that we can take a very large $z$-value that is bigger than both of the roots above such that the quadratic will have a positive value. Thus, the quadratic will have non-positive values only when $z$ is between both roots. So we must have
$$ \left|z - \frac{L_{x}^{2}}{2 \pi^{2}} \right| \le \frac{L_{x}^{2}}{2 \pi^{2}L_{y}}\sqrt{ L_{y}^{2} +  4(k \pi)^{2}},   $$
which means 
\begin{equation} \label{j_squared_bound}  
\frac{L_{x}^{2}}{2 \pi^{2}} - \frac{L_{x}^{2}}{2 \pi^{2}L_{y}}\sqrt{ L_{y}^{2} +  4(k \pi)^{2}}  \le j^{2}   \le \frac{L_{x}^{2}}{2 \pi^{2}} + \frac{L_{x}^{2}}{2 \pi^{2}L_{y}}\sqrt{ L_{y}^{2} +  4(k \pi)^{2}}. 
\end{equation}
Notice that $j$ is bounded above by 
$$|j| \le L_{x}\sqrt{\frac{1}{2 \pi^{2}} + 
\frac{\sqrt{ L_{y}^{2} +  4(k \pi)^{2}}}{2 \pi^{2}L_{y}}}. $$
On the other hand, since $\frac{\sqrt{ L_{y}^{2} +  4(k \pi)^{2}}}{L_{y}} \ge 1$, then the lower bound is always non-positive:
$$ \frac{L_{x}^{2}}{2 \pi^{2}} \left(1 - \frac{\sqrt{ L_{y}^{2} +  4(k \pi)^{2}}}{L_{y}} \right) \le 0, $$
which means we can neglect the lower bound in (\ref{j_squared_bound}).
\end{proof}
The bound (\ref{k_bound}) tells us that the lower bound of $|k|$ increases as $j$ increases. This is consistent with the red lines in Figure \ref{fig:trimming_comparison2}, on which $F_{2}(j,k)=0$. All combinations of $(j,k)$ in between the red lines are allowed. The red lines increase to the right or left, as $|j|$ increases. Moreover, the bound (\ref{k_bound}) also shows that increasing $L_{y}$ will increase the lower bound of $|k|$, which means excluding more $|k|$ values from below. As we can see from the Figure, the case $L_{x}=250,L_{y}=100$ noticeably allows more $k$ values than the case $L_{x}=250,L_{y}=200$. Similar behavior can be seen between the cases $L_{x}=150,L_{y}=100$ and $L_{x}=150,L_{y}=200$.

The bound (\ref{j_bound}) tells us that the range of allowed $j$ values increases as $|k|$ increases. In addition, increasing $L_{x}$ will allow more $j$ values in combination with lower $k$ values. This is consistent with Figure \ref{fig:trimming_comparison2}. The plots on the first row of Figure \ref{fig:trimming_comparison2} (with $L_{x}=250$) has wider range of accepted $j$ values paired with lower $k$ values than the plots on the second row (with $L_{x}=150$). 

\section{SCHEME 1: RUNGE-KUTTA AND PSEUDO-SPECTRAL} \label{sec:numerical_scheme}
\noindent We now present the pseudo-spectral numerical scheme that utilizes the bound (\ref{stable_condition_2}). We can write (\ref{2d_bbe}) as a first-order system by letting $v = u_{t}$ as follows
\begin{equation}\label{system}
\begin{bmatrix}
u_{t} \\ v_{t}
\end{bmatrix} = 
\begin{bmatrix}
v \\
u_{xx} + u_{xxxx} + u_{yy} -3 (u^{2})_{xx}.
\end{bmatrix}
\end{equation}
We view the solutions $u,v$ inside the domain $(-L_{x},L_{x}) \times (-L_{y},L_{y})$ using Fourier series. Let $x_{n_{1}}=-L_{x} + (n_{1}-1) \triangle x$, $y_{n_{2}}=-L_{y} + (n_{2}-1) \triangle y$, for $n_{1}=1,...,N_{1}$ and $n_{2}=1,...,N_{2}$, where $\triangle x = \frac{2L_{x}}{N_{1}-1}$ and $\triangle y = \frac{2L_{y}}{N_{2}-1}$. Let $U,V$ be the approximate solutions of $u,v$ at each point $(x_{n_{1}},y_{n_{2}},t)$. We write the approximate solution $U$ as follows

\begin{align}\label{u_approximate}
u(x,y,t) &= \sum _{k \in \mathbb{Z} } \sum_{j \in \mathbb{Z}} \widehat{u}_{j,k}(t)e^{i 2 \pi (\omega_{j}x + \omega_{k}y)}, \nonumber \\
U(x,y,t) &= \sum _{k \in \zeta_{2} } \sum_{j \in \zeta_{1}} \widehat{U}_{j,k}(t)e^{i 2 \pi (\omega_{j}x + \omega_{k}y)}, 
\end{align}
where $\omega_{j}=j/(2L_{x}), \: \omega_{k}=k/(2L_{y})$, and
\begin{equation} \label{wave_number1}
\zeta_{1} = \left( \zeta_{1,1}, \zeta_{1,2}, \hdots, \zeta_{1,N_{1}} \right) = \begin{cases}
(0,1,\hdots, \frac{N_{1}}{2}-1,-\frac{N_{1}}{2},\hdots,-1), \:\:\: \text{if }N_{1}\text{ is even} \\
(0,1,\hdots, \frac{N_{1}-1}{2}, -\frac{N_{1}-1}{2},\hdots,-1),  \:\:\: \text{if }N_{1}\text{ is odd}.\\
\end{cases} 
\end{equation}
\begin{equation} \label{wave_number2}
\zeta_{2} = \left( \zeta_{2,1}, \zeta_{2,2}, \hdots, \zeta_{2,N_{2}} \right) = \begin{cases}
(0,1,\hdots, \frac{N_{2}}{2}-1,-\frac{N_{2}}{2},\hdots,-1), \:\:\: \text{if }N_{2}\text{ is even} \\
(0,1,\hdots, \frac{N_{2}-1}{2}, -\frac{N_{2}-1}{2},\hdots,-1),  \:\:\: \text{if }N_{2}\text{ is odd}.\\
\end{cases} 
\end{equation}
Let $w=u^{2}$, we write $w$ as a different Fourier series. The approximate solutions in Fourier series for $v$ and $w$ are
\begin{align*}
V(x,y,t) = U_{t}(x,y,t) = \sum _{k \in \zeta_{2} } \sum_{j \in \zeta_{1}} \widehat{U}_{j,k}'(t)e^{i 2 \pi (\omega_{j}x + \omega_{k}y)}, \\
W(x,y,t) = \sum _{k \in \zeta_{2} } \sum_{j \in \zeta_{1}} \widehat{W}_{j,k}(t)e^{i 2 \pi (\omega_{j}x + \omega_{k}y)}.
\end{align*}
By plugging $U,V,W$ to the system (\ref{system}), we get the following system of ordinary differential equations for $\forall (j,k) \in \zeta_{1} \times \zeta_{2}$
\begin{align} \label{system_fourier}
\widehat{U}_{j,k}' &= \widehat{V}_{j,k}, \\
\widehat{V}_{j,k}' &= [-(2\pi \omega_{j})^{2} +(2\pi \omega_{j})^{4} - (2\pi \omega_{k})^{2}]\widehat{U}_{j,k} + 3(2\pi \omega_{j})^{2}  \widehat{W}_{j,k}. \nonumber
\end{align}
For all $t \ge 0$, let $\textbf{U}(t),\textbf{V}(t),\textbf{W}(t)$ be matrices of size $N_{1} \times N_{2}$ with elements that are the values of functions $U,V,W$ at points $(x_{n_{1}},y_{n_{2}},t)$: the entry at row $a$ and column $b$ is the value of each function at point $(x_{a},y_{b},t)$. Additionally, let $\widehat{\textbf{U}}(t),\widehat{\textbf{V}}(t),\widehat{\textbf{W}}(t)$ be matrices of size $N_{1} \times N_{2}$ with elements that are the values of the Fourier coefficients at time $t$: the entry at row $a$ and column $b$ is the coefficient of the Fourier mode at wave-index $j=\zeta_{1,a}, k=\zeta_{2,b}$. Define FFT($\cdot$) as the operator of two-dimensional discrete Fourier transform. Define the `trimming' matrix $\mathcal{D}=\mathcal{D}(L_{x},L_{y})$, a matrix of size $N_{1} \times N_{2}$, such that the entry at row $a$, column $b$ is
\begin{equation}
    \mathcal{D}_{a,b}  = \begin{cases}
        0, \:\:\: \text{if } \frac{L_{x}^{4} \zeta_{2,b}^{2} + (L_{x}L_{y})^{2}\zeta_{1,a}^{2}}{L_{y}^{2}\zeta_{1,a}^{4}} < \pi^{2}, \\
        1, \:\:\: \text{if } \frac{L_{x}^{4} \zeta_{2,b}^{2} + (L_{x}L_{y})^{2}\zeta_{1,a}^{2}}{L_{y}^{2}\zeta_{1,a}^{4}} \ge \pi^{2}, \:\: \text{or if} \:\: (\zeta_{1,a},\zeta_{2,b})=(0,0).
    \end{cases}
\end{equation}
Note that every operation of FFT($\cdot$) must be followed by element-wise multiplication by $\mathcal{D}$ to ensure the `trimming' of high-frequency modes. The element-wise multiplication will cancel the high-frequency unstable modes and preserve the stable ones. At time $t=0$, we set $\widehat{\textbf{U}}(0)= \mathcal{D} \odot \text{FFT}(\textbf{U}(0))$ and $\widehat{\textbf{V}}(0)= \mathcal{D} \odot \text{FFT}(\textbf{V}(0))$, where $\odot$ represents element-wise multiplication. We then solve (\ref{system_fourier}) numerically using the Runge-Kutta fourth order \cite{rk_4} method in Octave software \cite{octave}, manually. At each time step, we update (\ref{system_fourier}) with $\widehat{\textbf{W}}$ computed as
\begin{equation}\label{nonlinear_term}
    \widehat{\textbf{W}}(t) = \mathcal{D} \odot \text{FFT} \left( \left[\text{FFT}^{-1} ( \widehat{\textbf{U}}(t))\right]^{2} \right).
\end{equation}
After each time iteration, we take the inverse at each time point $t_{m}=m \triangle t$ to retain the solutions $U$ and $V$:
\begin{equation}
\textbf{U}(t_{m}) = \text{FFT}^{-1}(\widehat{\textbf{U}}(t_{m})), \:\: \textbf{V}(t_{m}) = \text{FFT}^{-1}(\widehat{\textbf{V}}(t_{m})).
\end{equation}

\section{SCHEME 2: STRANG SPLITTING AND PSEUDO-SPECTRAL}
\noindent As an alternative scheme other than Runge-Kutta 4, we present the Strang Splitting method \cite{strang_splitting_good_boussinesq, tao_kdv, ems_book}. 
\subsection{Intro: Strang Splitting on ODE}
\noindent The Strang splitting is a method for solving a differential equation by first separating between the linear and nonlinear parts, before combining the solutions of each individual part in a particular way. As a warm-up example, let us consider nonlinear ODEs of the form:
$$ y'(t) = A(y) + B(y), \:\:\: y(0) = y_{0}, $$
where $A$ and $B$ are linear and nonlinear functions of $y$, respectively. To use Strang splitting method, first we must have solutions, or approximate solutions in case of numerics, for each of the following linear and nonlinear equations with arbitrary initial values:
\begin{align}\label{ode_linear}
y'(t) = A(y), \:\:\: y(0) = y_{0},
\end{align} 
and
\begin{align}\label{ode_nonlinear}
y'(t) = B(y), \:\:\: y(0) = y_{0}.
\end{align} 
Let the operation $\phi_{A}(t) \circ y_{0}$ represents the solution of the linear equation (\ref{ode_linear}) with variable initial value $y_{0}$. Similarly, let the operation $\phi_{B}(t) \circ y_{0}$ represents the solution of the nonlinear equation (\ref{ode_nonlinear}) with variable initial value $y_{0}$. Let the time be discretized as $t_{m}=m \Delta t$, $m = 0,1,\hdots, M$. Let $y(t_{m})$ is known for some $m$, then we can approximate $y(t_{m+1})$ via Strang splitting method. The Strang splitting method has three steps. In the 1st step, we compute the nonlinear solution of (\ref{ode_nonlinear}) at time $t=\Delta t/2$ with $y(t_{m})$ as initial value: we compute 
\begin{equation}\label{step_1}
\phi_{B}(\Delta t/2) \circ y(t_{m}).
\end{equation}
In the 2nd step, we use (\ref{step_1}) as the initial value for the linear solution of (\ref{ode_linear}) and compute the solution at time $t=\Delta t$: we compute
\begin{equation}\label{step_2}
\phi_{A}(\Delta t) \circ \left( \phi_{B}(\Delta t/2) \circ y(t_{m}) \right).  
\end{equation}
Lastly, in the 3rd step we use (\ref{step_2}) as the initial value again for the nonlinear solution of (\ref{ode_nonlinear}) and compute the solution at time $t=\Delta t/2$: we compute
$$  \phi(\Delta t/2) \circ \left( \phi_{A}(\Delta t) \circ \left( \phi_{B}(\Delta t/2) \circ y(t_{m}) \right) \right).  $$
This last value is considered as approximation for $y(t_{m+1})$, that is
$$ y(t_{m+1}) \approx \phi_{B}(\Delta t/2) \circ \phi_{A}(\Delta t) \circ \phi_{B}(\Delta t/2) \circ  y(t_{m}). $$

\begin{figure}[H]
\centering
\includegraphics[width=12cm]{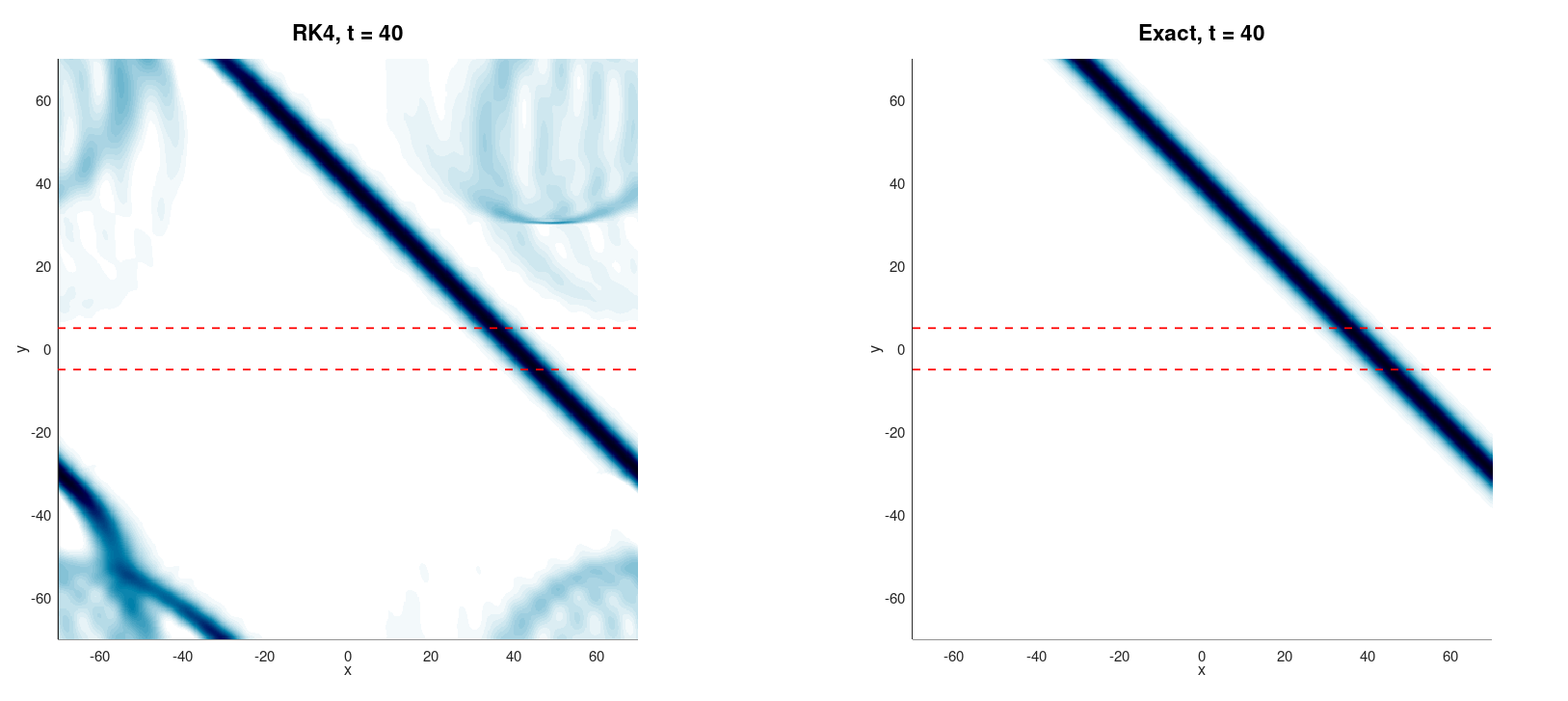}
\caption{The colormap plot of the numerical solution at $t=40$ for the first simulation via Runge-Kutta method (left). The red dashed lines indicate $y=-5$ and $y=5$, we only compute the error in between these lines.}
\label{fig:rk4_sim1}
\end{figure}

\subsection{Strang Splitting on ``bad'' Boussinesq Equation with Pseudo Spectral Fourier}
\noindent Moving on from ODE, the Strang splitting method can also be used to numerically solve the two-dimensional ``bad'' Boussinesq system (\ref{system}). Here, we use the same notations for Fourier series and matrices mentioned in Section \ref{sec:numerical_scheme}. The system (\ref{system}) can be viewed as 
\begin{equation}
Y_{t} = A(Y) + B(Y), \:\:\: Y(x,y,t) = [u(x,y,t), v(x,y,t)]^{T},
\end{equation}
where
$$ A(Y) = \begin{bmatrix} 
v\\
u_{xx} + u_{xxxx} + u_{yy} 
\end{bmatrix}, \:\:\: B(Y) = \begin{bmatrix} 0 \\ -3 (u^{2})_{xx}   \end{bmatrix}.$$
Let the operation $\Phi_{A}(x,y,t) \circ Y_{0}$ represents the solution (whether exact or numerical) of $Y_{t}=A(Y)$ with $Y_{0}$ as the variable for the initial condition vector. Let the operation $\Phi_{B}(x,y,t) \circ Y_{0}$ represents the solution (whether exact or numerical) of $Y_{t}=B(Y)$ with $Y_{0}$ as the variable for the initial condition vector. Similar to the ODE case, the Strang splitting scheme is
\begin{align} \label{STRANG_scheme}
&[u(x,y,t_{m+1}),v(x,y,t_{m+1})]^{T} \\
&\approx  \Phi_{B}(x,y,\Delta t/2) \circ \Phi_{A}(x,y,\Delta t) \circ \Phi_{B}(x,y,\Delta t/2) \circ [u(x,y,t_{m}), v(x,y,t_{m})]^{T}. \nonumber
\end{align}
To be able to use (\ref{STRANG_scheme}), we need to know at least how to approximate $\Phi_{A}$ and $\Phi_{B}$. We start with the linear system:
\begin{align} \label{STRANG_A}
u_{t} &= v, \\
v_{t} &= u_{xx} + u_{xxxx} + u_{yy},
\end{align}
with variable initial condition vector $Y_{0}=[u_{0},v_{0}]^{T}$. This linear system is exactly the linearized two-dimensional Boussinesq equation (\ref{2d_bbe_linear}), and we can solve it in the frequency space with exact solution (\ref{uhat_exact_solution}). Numerically, using pseudo-spectral Fourier, we approximate (\ref{uhat_exact_solution}) for each 

\begin{figure}[H]
\centering
\includegraphics[width=12cm]{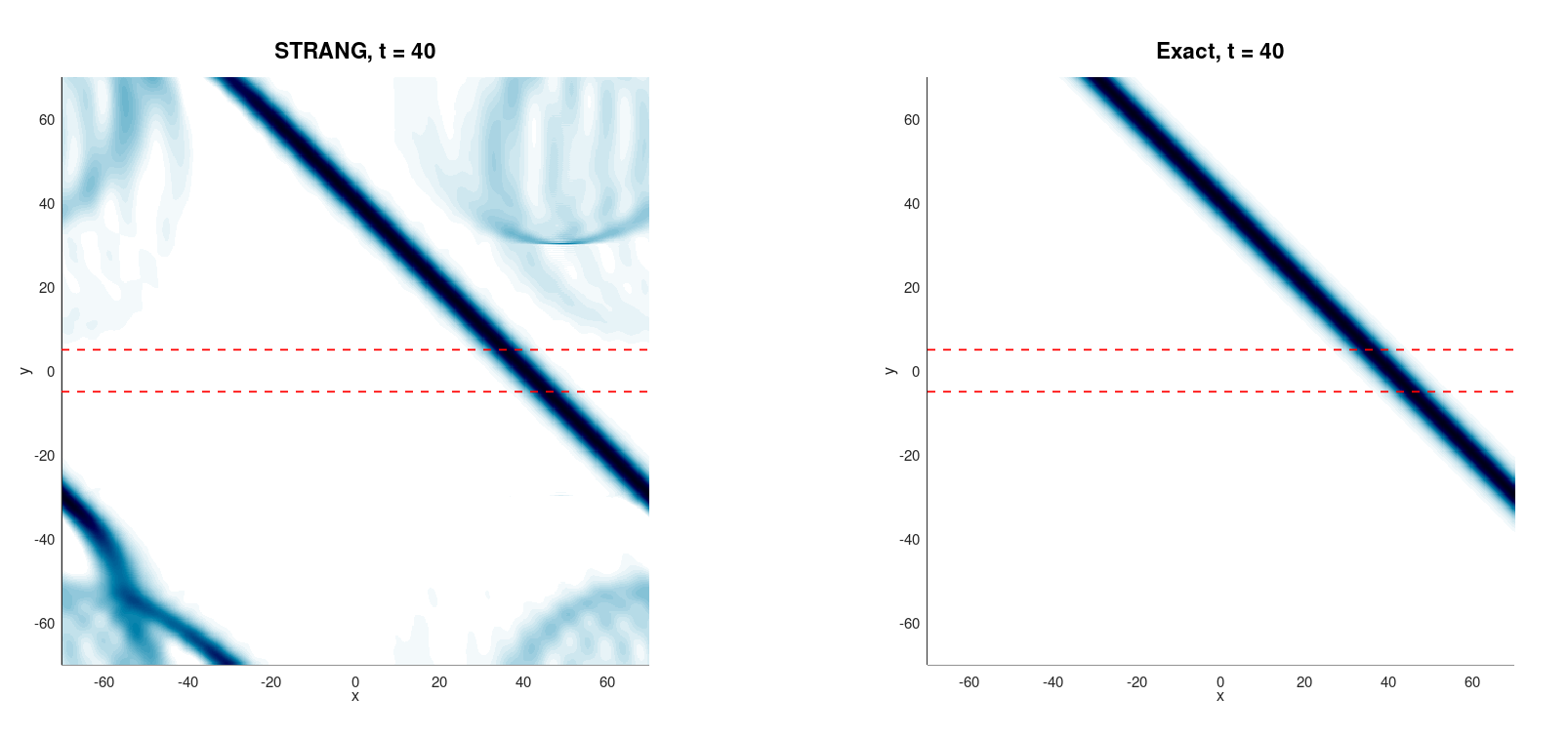}
\caption{The colormap plot of the numerical solution at $t=40$ for the first simulation via Strang splitting method (right). The red dashed lines indicate $y=-5$ and $y=5$, we only compute the error in between these lines.}
\label{fig:strang_sim1}
\end{figure}

\noindent point $(j,k)$ as:
\begin{align} \label{exact_linear_STRANG}
\widehat{U}_{j,k}(t) &= 
\begin{cases} 
\widehat{V}_{j,k}(0)t + \widehat{U}_{j,k}(0), \:\:\: \lambda_{j,k} = 0 \\
  \frac{\widehat{U}_{j,k}(0)+ \lambda_{j,k}^{-1/2}\widehat{V}_{j,k}(0)}{2} e^{\sqrt{\lambda_{j,k}} t} + \left( \widehat{U}_{j,k}(0) - \frac{\widehat{U}_{j,k}(0)+\lambda_{j,k}^{-1/2}\widehat{V}_{j,k}(0)}{2} \right) e^{-\sqrt{\lambda_{j,k}} t}, \:\:\: \lambda_{j,k} > 0 \\
\frac{\widehat{U}_{j,k}(0)+ \frac{\widehat{V}_{j,k}(0)}{ i\sqrt{-\lambda_{j,k}}}}{2} e^{i\sqrt{-\lambda_{j,k}} t} +  \left( \widehat{U}_{j,k}(0) - \frac{\widehat{U}_{j,k}(0)+\frac{\widehat{V}_{j,k}(0)}{ i\sqrt{-\lambda_{j,k}}}}{2} \right)e^{-i\sqrt{-\lambda_{j,k}} t}  , \:\:\: \lambda_{j,k} < 0,
\end{cases} \\
\widehat{V}_{j,k}(t) &= \widehat{U}_{j,k}'(t). \nonumber
\end{align}
We will only use the case where $\lambda_{j,k} \le 0$, before performing inverse discrete Fourier transform. The inverse discrete Fourier transform of $\mathcal{D} \odot \widehat{\textbf{U}}(t)$ and $\mathcal{D} \odot \widehat{\textbf{V}}(t)$ computed from (\ref{exact_linear_STRANG}) are the approximation for $\Phi_{A}(x,y,t) \circ Y_{0}$ on all discrete spatial points $(x_{n_{1}},y_{n_{2}})$. 

Next, we will approximate $\Phi_{B}$ for the nonlinear system:
\begin{align} \label{STRANG_B}
u_{t} &= 0, \\
v_{t} &= -3 (u^{2})_{xx},
\end{align}
with variable initial condition vector $Y_{0}=[u_{0},v_{0}]^{T}$. Via pseudo-spectral Fourier, the equivalent system in frequency space for each point $(j,k)$ can be approximated as:
\begin{align} \label{STRANG_B_frequency}
\widehat{U}_{j,k}' &= 0 \\
\widehat{V}_{j,k}' &= 3(2\pi \omega_{j})^{2}  \widehat{W}_{j,k}. \nonumber
\end{align} 
Since the time derivative for the first variable is 0, we have constant $\widehat{U}_{j,k}(t) = \widehat{U}_{j,k}(0)$ for each $(j,k)$. As a consequence, we have $ \widehat{\textbf{W}}(t) = \text{FFT} \left( [\text{FFT}^{-1}(\widehat{\textbf{U}}(0))]^{2} \right) $, which means $\widehat{W}_{j,k}(t)$ is constant and therefore $\widehat{V}_{j,k}(t)$ is linear. Therefore, the solution for (\ref{STRANG_B_frequency}) can be summarized as
\begin{align} \label{exact_nonlinear_STRANG}
\widehat{\textbf{U}}(t) &= \widehat{\textbf{U}}(0), \\
\widehat{\textbf{V}}(t) &= \widehat{\textbf{V}}(0) + t \left( 3(2\pi \vec{\omega}_{x})^{2} \odot \text{FFT} \left( [\text{FFT}^{-1}(\widehat{\textbf{U}}(0))]^{2} \right) \right), \nonumber
\end{align}
where $\vec{\omega}_{x}$ is a matrix with each column vector equals $[\omega_{\zeta_{1}}, \hdots, \omega_{\zeta_{N_{1}}}]^{T}$ and $\odot$ is element-wise multiplication. Again, we only consider (\ref{exact_nonlinear_STRANG}) for the points $(j,k)$ where $\lambda_{j,k} \le 0$. We then 

\begin{figure}[H]
\centering
\includegraphics[width=8cm]{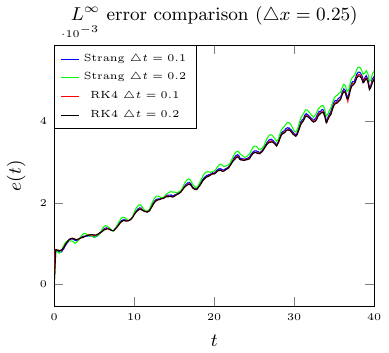}
\caption{The $L^{\infty}$ error for both numerical methods up to $t=40$ for the first simulation, using $\triangle x=0.25$ and two different time steps $\Delta t=0.1,0.2$.}
\label{fig:error_sim1}
\end{figure}

\noindent perform inverse discrete Fourier transform of $\mathcal{D} \odot \widehat{\textbf{U}}(t)$ and $\mathcal{D} \odot \widehat{\textbf{V}}(t)$ to get the approximation for $\Phi_{B}(x,y,t) \circ Y_{0}$ on all discrete spatial points $(x_{n_{1}},y_{n_{2}})$.

\section{NUMERICAL RESULTS}
\noindent The numerical computation is conducted using Octave software and especially its built-in two-dimensional fast Fourier transform function \texttt{fft2} and the inverse \texttt{ifft2}. The exact solution of (\ref{2d_bbe}) from \cite{exact_barrera} is of the form
\begin{equation}\label{exact_sol_raw}
u(x,y,t) = - \frac{2 \alpha^{4} b_{0} e^{\alpha x +\beta y - t\sqrt{\beta^{2} + \alpha^{2} + \alpha^{4}}}}{(1 + \alpha^{2}b_{0}e^{\alpha x + \beta y - t \sqrt{\beta^{2} + \alpha^{2} + \alpha^{4}}} )^{2}}.
\end{equation}
This solution is a soliton wave travelling at constant speed. Since the exact solution is a travelling soliton, then it is mass conserving, so we do not need to check whether $\int_{\Omega} u_{t}(x,y,0) \: d\Omega = 0$. To measure accuracy at each time point $t_{m}$, we use the $L^{\infty}$ error function
\begin{equation*}
    e(t) =   \sup_{(x,y) \in \tilde{\Omega}} \left|U(x,y, t)-u(x,y, t) \right|,
\end{equation*}
where $\tilde{\Omega} = \{ x_{n_{1}}, \:\: 1 \le n_{1} \le N_{1} \} \times \{ y_{n_{2}}, \:\: 1 \le n_{2} \le N_{2} \}$. 

For the first setting of initial condition and domain, we use initial condition of the form (\ref{exact_sol_raw}) with $\alpha=0.6$, $\beta=0.6$, and $b_{0}=10^{6}$ over the domain $\Omega = [-70,70]^{2}$. The exact solution with this setting is a soliton wave heading north-east direction with constant speed. Using this setting, we numerically solve the Boussinesq system (\ref{system}) for $t \in [0,40]$ using both the Runge-Kutta method and Strang splitting with $\Delta x=\Delta y=0.25$ and $\Delta t = 0.1,0.2$. Due to the automatic Fourier periodic boundary condition, waves passing through the boundary will reappear again from the opposite side. As a consequence, we only compute the $L^{\infty}$ error over the subdomain $y \in [-5,5]$: in this subdomain, the waves from periodic boundary condition will not appear at least before $t=40$. The plot of the numerical solutions using $\Delta x=\Delta y=0.25$ and $\Delta t=0.1$ are shown in Figure \ref{fig:rk4_sim1} and \ref{fig:strang_sim1}. As we can see, the effect of periodic boundary condition can be seen around the corners of the domain. However, both numerical solutions have solitons heading in the same direction as the exact solution, and with approximately the same shape as well. The $L^{\infty}$ error is plotted in Figure \ref{fig:error_sim1} for both time step scenarios $\Delta t=0.1,0.2$. We can see 

\begin{figure}[H]
\centering
\includegraphics[width=12cm]{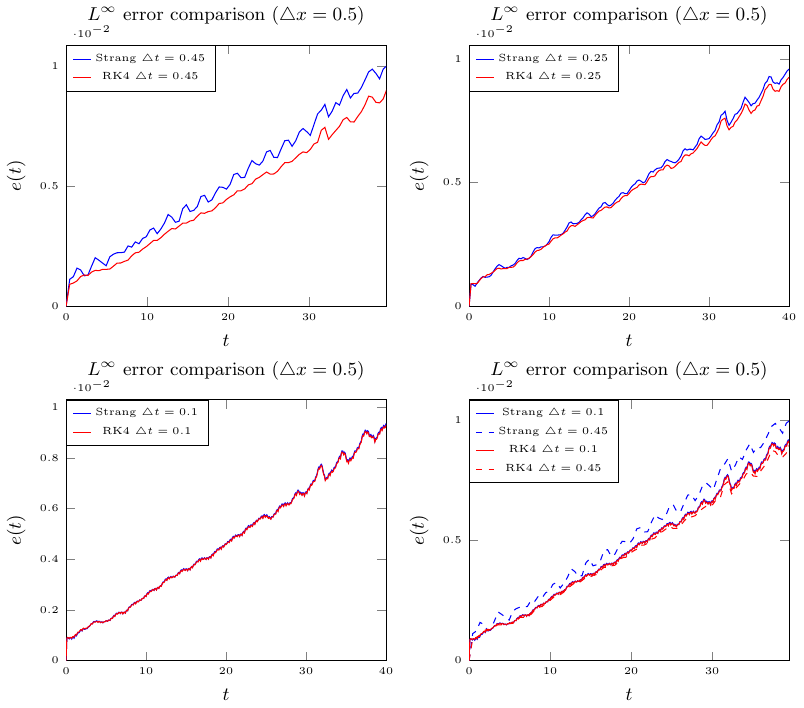}
\caption{The $L^{\infty}$ error for both numerical methods up to $t=40$ for the first simulation, using $\triangle x=0.5$ and three different time steps $\Delta t=0.1,0.25,0.45$.}
\label{fig:error_sim2}
\end{figure}

\noindent that both methods produce good results, with $e(40) \approx 0,005 < 2.8 \%$ of the initial condition's absolute peak $\max|U(x,y,0)|=0.18$. The error plot shows that the performance of Runge-Kutta 4 (RK4) is slightly better than Strang splitting for $t \in [5,40]$ for $\Delta t = 0.2$. However, for $\Delta t=0.1$, the performance of both methods are almost indistinguishable. Additionally, we also simulate with $\Delta x=\Delta y=0.5$ and $\Delta t = 0.1,0.25,0.45$ to compare the numerical errors. The $L^{\infty}$ errors resulting from cases $\Delta x=\Delta y = 0.5$ with $\Delta t = 0.1,0.25,0.45$ are shown in Figure \ref{fig:error_sim2}. From these plots, it is clear that the performance difference between Runge-Kutta 4 and Strang splitting becomes more noticeable as $\Delta t$ increases. For $\Delta t=0.1,0.25,0.45$, the Runge-Kutta 4 performs well and shows similar results for all three time steps, while the Strang splitting becomes approximately as good as Runge-Kutta only when $\Delta = 0.1$.

\begin{figure}[H]
\centering
\includegraphics[width=14cm]{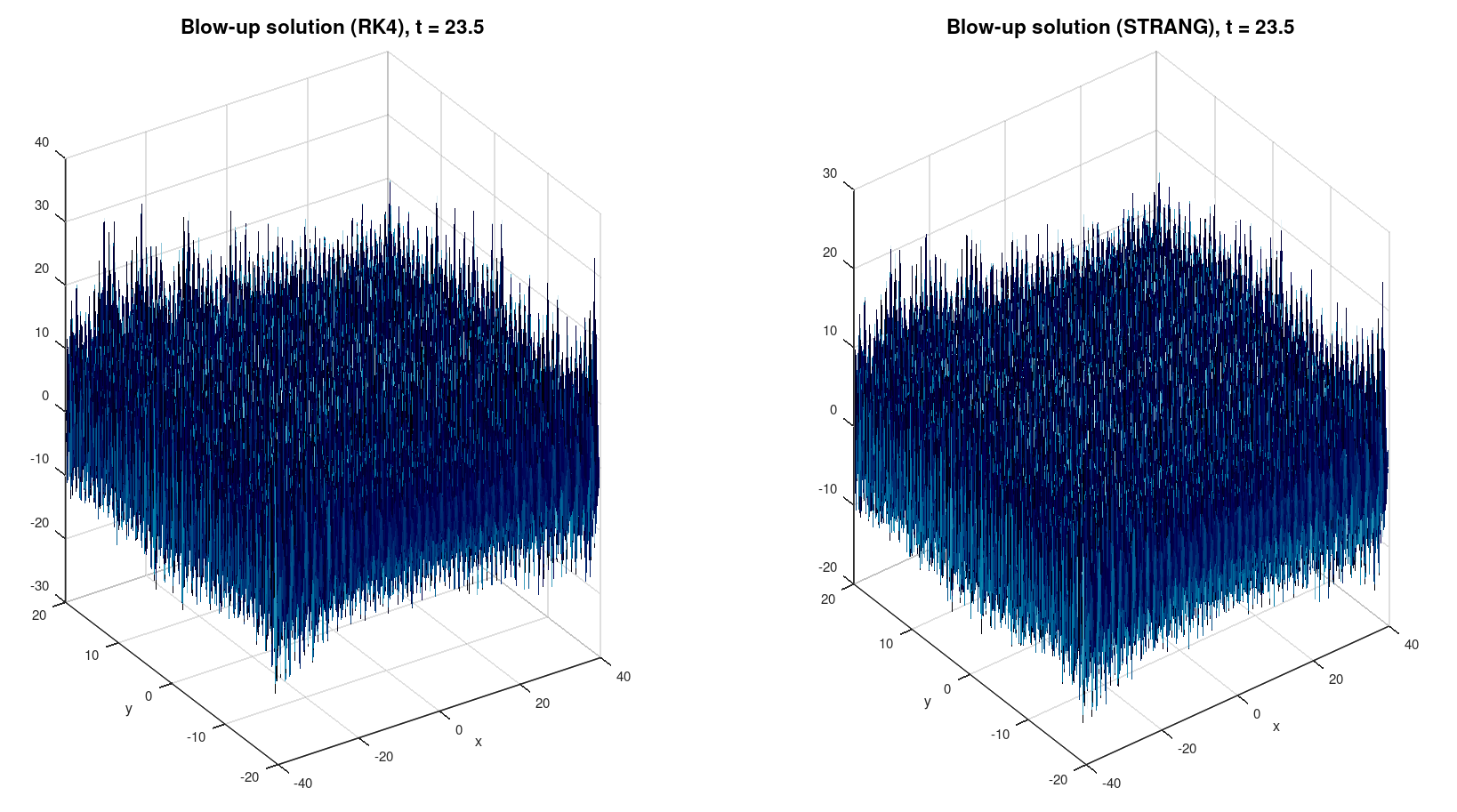}
\caption{Surface plots of the blow-up numerical solutions at $t=23.5$ when we minorly violate the stable condition (\ref{stable_condition_2}).}
\label{fig:blowup_example}
\end{figure}

\begin{figure}[H]
\centering
\includegraphics[width=14cm]{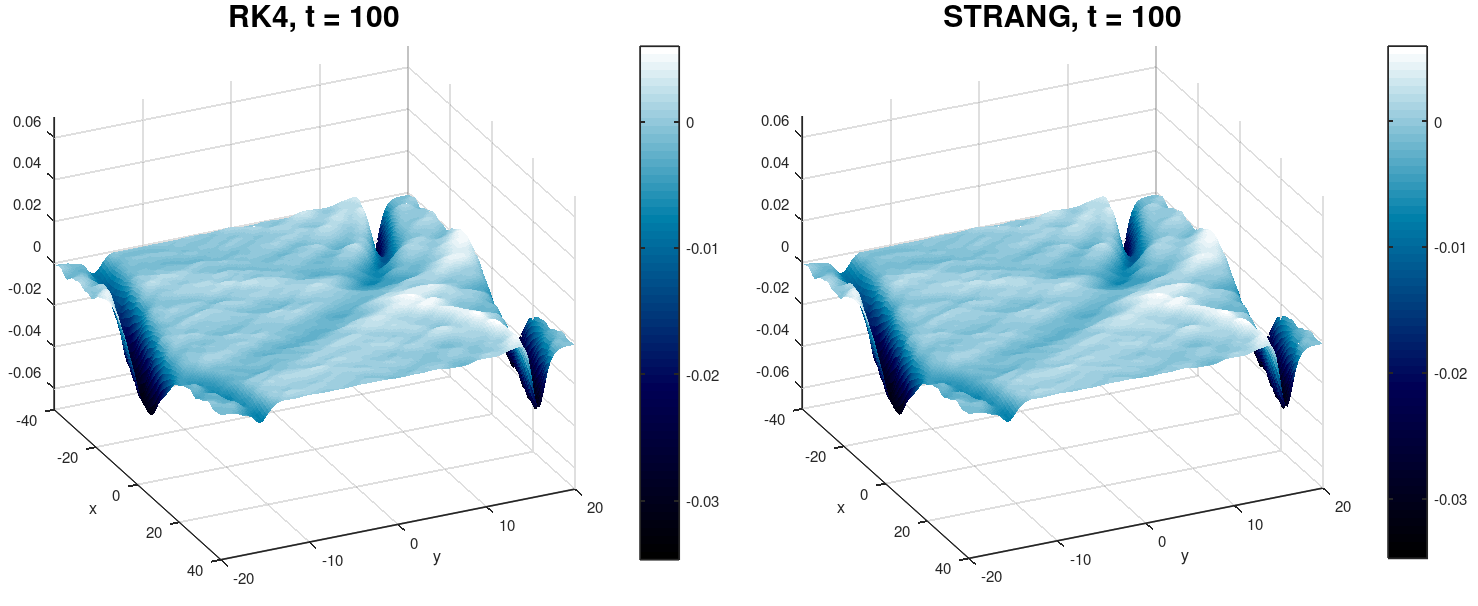}
\caption{Surface plots of the numerical solutions at $t=100$ via Runge-Kutta and Strang Splitting methods by following the stable condition (\ref{stable_condition_2}).}
\label{fig:long_sim_3d}
\end{figure}

For the second simulation, we test whether our numerical scheme really depends on the linear `trimming' condition (\ref{stable_condition_2}). We tried a simulation where we minorly disobey the `trimming' condition: we include the points $(j,k)$ where 
$$ 0 \le L_{x}^{4} k^{2} + (L_{x}L_{y})^{2}j^{2} - 0.99\pi^{2}L_{y}^{2}j^{4},   $$
therefore allowing a few of the `illegal' high-frequency Fourier modes in our solution. We use initial condition (\ref{exact_sol_raw}) with $\alpha=0.25$, $\beta=2$, and $b_{0}=16$, over the domain $(x,y) \in [-40,40] \times [-20,20]$. The discrete spatial and time steps are $\Delta x=\Delta y=0.25$ and $\Delta t =0.1$. After performing the numerical simulation, it turns out that at time $t=23.5$, a blow-up solution is apparent as shown in Figure \ref{fig:blowup_example}. However, we also did a longer simulation using the same setting but obeying the linear `trimming' condition (\ref{stable_condition_2}), and the results are stable solutions up to $t=100$ (see Figure \ref{fig:long_sim_3d}). This suggests that the nonlinear Boussinesq equation (\ref{2d_bbe}) also depends on the linear condition (\ref{stable_condition_2}).

\begin{figure}[H]
\centering
\includegraphics[width=9cm]{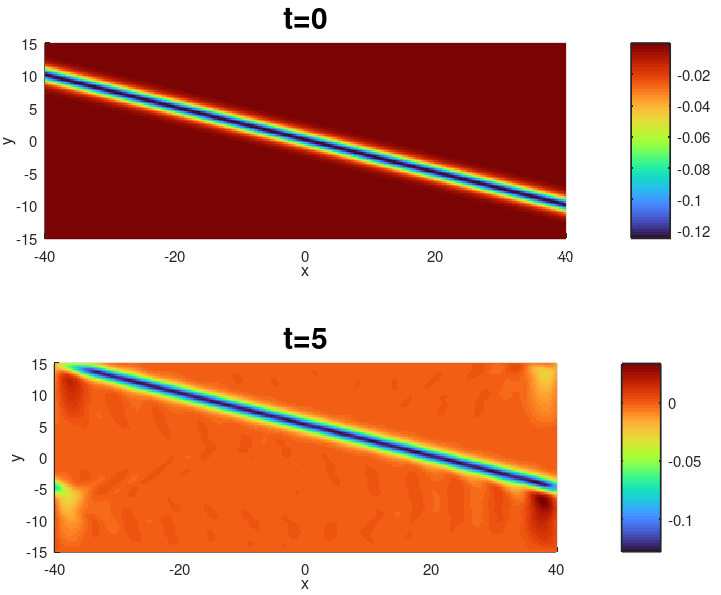}
\caption{Numerical solution at time $t=0$, $t=5$ with Dirichlet boundary condition.}
\label{fig:rectangular_1}
\end{figure}

\begin{figure}[H]
\centering
\includegraphics[width=9cm]{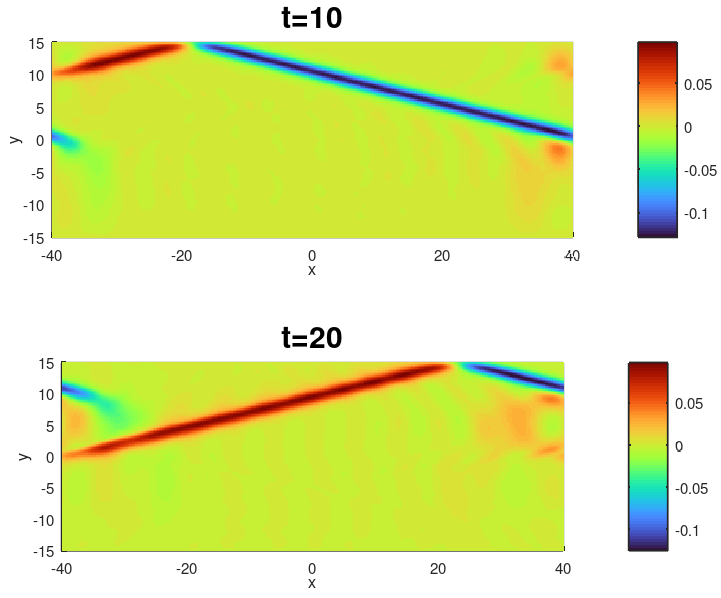}
\caption{Numerical solution at time $t=10$, $t=20$ with Dirichlet boundary condition.}
\label{fig:rectangular_2}
\end{figure}

\begin{figure}[H]
\centering
\includegraphics[width=9cm]{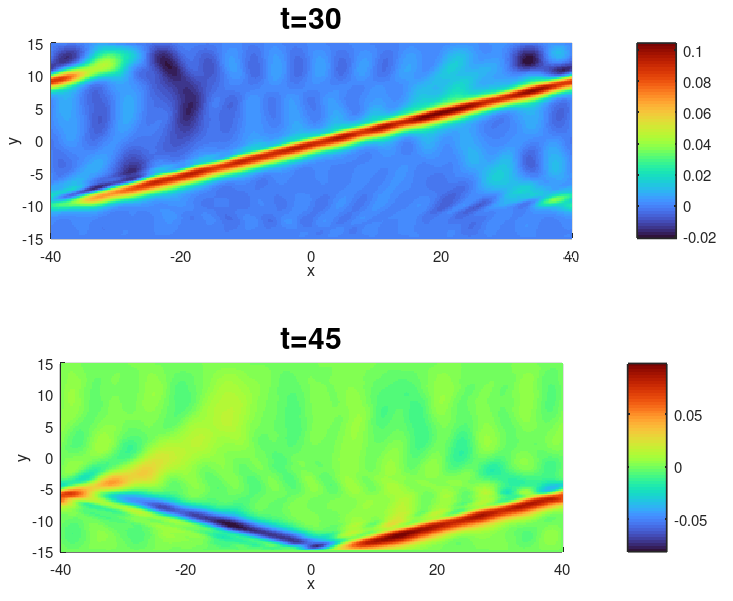}
\caption{Numerical solution at time $t=30$, $t=45$ with Dirichlet boundary condition.}
\label{fig:rectangular_3}
\end{figure}

Lastly, we perform another numerical simulation on a rectangular domain $[-40,40] \times [-20,20]$ with Dirichlet boundary condition $U(x,y,t)=0$ on the set of spatial points
\begin{equation*} 
\partial \Omega_{\text{dirichlet}} = \{ (x,y) | x \in [-40, -40], y \in [-20,-15] \} \cup \{ (x,y) | x \in [-40, 40], y \in [15,20] \}. 
\end{equation*}
To implement the Dirichlet boundary condition, we define a boundary matrix \textbf{B} of the same size as \textbf{U}. Matrix \textbf{B} is such that all the matrix elements corresponding to spatial points $(x,y) \notin \partial \Omega_{\text{dirichlet}}$ have values 1, while the others have value $0$. In each time iteration after we inverse the $\widehat{\textbf{U}}$ and $\widehat{\textbf{V}}$ to get the solution \textbf{U} and \textbf{V}, we perform element-wise multiplication $\textbf{B} \odot \textbf{U}$ and $\textbf{B} \odot \textbf{V}$, and then Fourier transform again to get the new $\widehat{\textbf{U}}$ and $\widehat{\textbf{V}}$ for Dirichlet boundary condition. The results are shown in Figure \ref{fig:rectangular_1}-\ref{fig:rectangular_3}. We can see a part of the soliton has been reflected against the north wall at $t=10$ indicated by red color, and at $t=20$ longer part has been reflected. At $t=30$, we can see the approximate soliton is now heading south-east, and at $t=45$ we can see a part of it has been reflected off the south wall indicated by blue color.

\vspace{2cc}

\bibliographystyle{ieeetr}
\bibliography{references}

\end{document}